\documentclass[showpacs,preprintnumbers,amsmath,amssymb]{revtex4}
\newtheorem{theorem}{Theorem}

\newtheorem{lemma}{Lemma}
\newtheorem{corollary}{Corollary}

\newtheorem{condition}{Condition}
\newtheorem{example}{Example}
\newcommand{\BlackBox}{\rule{1.5ex}{1.5ex}}  
\newenvironment{proof}{\par\noindent{\bf Proof:\
}}{\hfill\BlackBox\\[2mm]}

\usepackage{graphicx}
\usepackage{dcolumn}
\usepackage{bm}

\begin{document}
\title{
Time-Energy Uncertainty Relation for a Quantum Clock as a Control
Device
}

\author{Takayuki Miyadera}
\affiliation{%
Research Center for Information Security, \\
National Institute of Advanced Industrial Science and Technology \\
1-1-1 Umezono, Tsukuba, 
Ibaraki 305-8561 Japan. 
\\
(E-mail: miyadera-takayuki@aist.go.jp)
}%


\date{\today}

\begin{abstract}
A quantum clock working as a control device is examined. 
The quality of the control process is characterized by the magnitude of deviation of perturbed state from unperturbed state of the controlled system. 
Uncertainty relations that relate the time duration of the process and 
energy of the clock to  
the quality of the control are presented. 
\end{abstract}
\maketitle
\section{Introduction}
We read a clock to know what time it is. While time is a classical external parameter 
in the   
equation of motion, we need to observe a physical object called a clock to know the time 
of occurrence of a certain event. 
In quantum theory, dynamics of any physical object, including a clock,  
is governed by the Schr\"odinger equation. 
Following the pioneering work by Salecker and Wigner \cite{Wigner}, several versions of the quantum clock have been proposed and analyzed 
\cite{Peres, Hilgevoord2}. 
In particular, properties of quantum clocks have been investigated in the context of 
the time-energy uncertainty relation \cite{Hilgevoord1,BuschTEUR,Busch1,AReznik,CReznik}. 
\par
In addition to basic roles, 
Peres \cite{Peres} argued that a clock can be employed to control the duration of perturbation on a system. 
Such control dynamics is often described by introducing a
time-dependent perturbation. For instance, the Hamiltonian of the system $H_s$ is 
replaced by a time-dependent term such as $H_s + f(t) B$, where 
$f(t)$ is a time-dependent classical quantity and $B$ is a self-adjoint operator. 
While there have been a lot of important results using this setting thus far,  
it is necessary to validate 
such a phenomenological treatment from a fundamental perspective. 
In fact, from a strict viewpoint, the 
Hamiltonian formalism, whose origins are in time uniformity, is applicable 
only to closed systems. 
Thus, a composite system consisting of the system and 
clock should be treated. 
Using a time-independent Hamiltonian for such a composite system, 
Peres \cite{Peres} 
presented 
a concrete model of a quantum clock that switches on a magnetic field to control the 
precession of a spinning particle. 
\par
In this paper, we attempt to define a quantum clock working as a control device.
In our definition, the clock evolves freely until it gives rise to a perturbation on the system. 
The quality of the control process is characterized by the magnitude of deviation of 
perturbed state from unperturbed state of the system. 
We show that the quality is related to the energy of the clock and 
perturbation time duration. 
\section{quantum clock as controlling device}
We consider two systems: One is a system to be controlled 
and other is a quantum clock that works as 
a timing device and controls the other. 
We denote the Hilbert space 
describing the system (resp. the clock) 
 by ${\cal H}_s$ (resp. ${\cal H}_c$). The own dynamics of the system (resp. clock) is governed by the 
Hamiltonian $H_s$ (resp. $H_c$) defined on ${\cal H}_s$ (resp. ${\cal H}_c$). 
\par
A clock is specified by the Hilbert space ${\cal H}_c$, 
Hamiltonian $H_c$ and a state $\phi_c(0) \in {\cal H}_c$ at time $t=0$, where 
$t$ represents an external time that appears in the Schr\"odinger equation as a parameter.  
Suppose that the clock is isolated. 
At time $t=0$ 
the state of the 
clock is $\phi_c(0)$. 
Thus for an arbitrary time $t \in {\bf R}$, the state at time $t$ is 
written as $\phi_c(t):=e^{-i \frac{H_c t}{\hbar}}\phi_c(0)$ 
\cite{mixed}. 
While there are several models of the quantum clock, 
we do not restrict ourselves to any concrete model. 
Instead, we require clocks to satisfy the following general conditions
that represent the capability to switch on a perturbation 
at a certain time. 
We require the clock to evolve freely 
up to time $t=0$ and  
switch on a perturbation at some time after $t=0$. 
We treat the system-plus-clock system as 
a closed system. 
An interaction between the 
clock and the system is denoted by $V$. Thus, the full Hamiltonian is   
\begin{eqnarray*}
H=H_s+H_c+V. 
\end{eqnarray*}
\par
At time $t\leq 0$, the perturbation has not been switched on. 
This condition is formulated as follows.
\begin{condition}\label{cond1} 
For any $t\leq 0$ and for any state $\Omega$ of 
the system, 
\begin{eqnarray}
V(\phi_c(t)\otimes \Omega)=0
\label{cond1eq}
\end{eqnarray}
holds. 
\end{condition}
Let us emphasize that in Condition \ref{cond1}, the left-hand side of (\ref{cond1eq}) 
must disappear for any  
$\Omega$. This condition implies 
that, only after the 
clock reaches $|\phi_c(0)\rangle$, $V$ makes the clock and the system 
interact with each other.  
This condition ensures that the composite system evolves independently 
up to time $t=0$. 
\begin{lemma}\label{lemma1}
If at a certain time $t_0 \leq 0$ the state has a product form
$\Theta(t_0)=|\phi_c(t_0\rangle 
\langle \phi_c(t_0)|\otimes \sigma_s(t_0)$; then  
the state $\Theta(t)$ at any time $t\leq 0$ becomes a product state  
\begin{eqnarray*}
\Theta(t)= |\phi_c(t)\rangle \langle \phi_c(t)| \otimes \sigma_s(t),  
\end{eqnarray*}
where 
$\sigma_s(t)=
e^{-i \frac{H_s}{\hbar}(t-t_0)} \sigma_s(t_0) e^{ i \frac{H_s}{\hbar} (t-t_0)}$.   
\end{lemma} 
\begin{proof}
Without loss of generality we may assume that 
$\sigma_s(t_0)$ is a pure state and is written 
as $\sigma_s(t_0)=|\Omega(t_0)\rangle \langle 
\Omega(t_0)|$. $\sigma_s(t)$ is written as 
$\sigma_s(t)=|\Omega(t)\rangle \langle \Omega(t)|$ with 
$|\Omega(t)\rangle =e^{-i \frac{H_s}{\hbar}(t-t_0)}|\Omega(t_0)\rangle$. 
It holds for $t\leq 0$ that 
\begin{eqnarray*}
i \hbar \frac{d}{dt}(\phi_c(t)\otimes \Omega(t))
&=&(H_c +H_s)\phi_c(t)\otimes \Omega(t)
\\
&=&
(H_c +H_s +V)\phi_c(t)\otimes \Omega(t)
=H (\phi_c(t)\otimes \Omega(t)). 
\end{eqnarray*}
where we used 
$V(\phi_c(t)\otimes \Omega(t))=0$. 
Integrating this equation, we obtain 
$\Theta(t)=|\phi_c(t)\rangle \langle \phi_c(t)|\otimes |\Omega(t)\rangle \langle \Omega(t)|$. 
\end{proof}
To avoid the trivial interaction $V=0$, we introduce the following condition.
\begin{condition}\label{cond2} 
There exists a state $\Omega\in {\cal H}_s$ and a time $t>0$ such that 
\begin{eqnarray*}
V e^{-i\frac{H}{\hbar}t}( \phi_c(0)\otimes \Omega ) \neq 0. 
\end{eqnarray*}
\end{condition}
Note that this condition is rather weak.  
It still allows the existence of 
$\Omega$ on which $V$ acts trivially for any $t\in {\bf R}$. 
In addition, this condition does not specify the switching time of the perturbation exactly 
as $t=0$. It only requires the 
existence of the moment at which the perturbation does not vanish. 
\par
The following example will be revisited. 
\begin{example}\label{ex1}
Suppose that a clock is described by 
${\cal H}_c=L^2({\bf R})$ and 
$H_c =p:=-i \hbar \frac{d}{dx}$. 
It is coupled with a system described by 
${\cal H}_s={\bf C}^2$ that has an orthonormalized basis $\{|0\rangle, |1\rangle\}$ and 
a trivial Hamiltonian 
$H_s=0$. 
The interaction term is introduced as 
$V=\int dx g(x) \otimes (|0\rangle \langle 0|-|1\rangle \langle 1|)$, where 
$g$ is a nonvanishing real function 
whose support is included in $(0, \Delta )$ for some $\Delta >0$. 
In the position representation $\phi_c(0)$ is supposed to be strictly localized in 
the negative real line.  
That is,  support of $\phi_c(t=0, x)$ is included in 
$(-\delta,0)$ for some $\delta>0$. 
The freely evolved state of the clock can be written as 
$\phi_c(t,x)=\phi_c(0,x-t)$, which has a support in $(-\delta+t, t)$. 
It is easy to see that this clock-plus-system satisfies Condition \ref{cond1}. 
\par
Let us consider the time evolution of the state 
denoted by 
$|\phi_c(0)\rangle \otimes |\Omega_s(0)\rangle$. 
For $|\Omega_s(0)\rangle =|0\rangle $, 
while the state of the system remains $|0\rangle$, 
the state of the clock suffers from the effect of the interaction. 
We denote the 
whole state at time $t$ 
 by $|\phi^0_c(t)\rangle \otimes |0\rangle$.   
It can be shown that at time $t$, the state of the clock 
$\phi^0_c(t)$ becomes 
$\phi^0_c(t,x)=e^{\frac{i}{\hbar}\int^x_0 dx' g(x')}\phi_c(t,x)$ in the position 
representation. 
On the other hand, 
for $|\Omega_s(0)\rangle =|1\rangle $, 
while the state of the system remains $|1\rangle$, 
the state $\phi^1_c(t)$ of the clock at time $t$  
becomes 
$\phi^1_c(t,x)=e^{-\frac{i}{\hbar}\int^x_0 dx' g(x')}\phi_c(t,x)$ in the position 
representation. 
Thus, for $|\Omega_s(0)\rangle =\frac{1}{\sqrt{2}}(|0\rangle +|1\rangle)$, 
the state of the system evolves as  
\begin{eqnarray*}
\rho_s(t)=\frac{1}{2}\left(
|0\rangle \langle 0|+|1\rangle \langle 1|
+|0\rangle \langle 1 | (\phi^1_c(t),\phi^0_c(t))
+|1\rangle \langle 0 |(\phi^0_c(t),\phi^1_c(t))
\right), 
\end{eqnarray*}
which does not agree with the freely evolved state $\rho^0_s(t)=\frac{1}{2}
\left(|0\rangle \langle 0|+|1\rangle \langle 1|
+|0\rangle \langle 1|+|1\rangle \langle 0|
\right)$ for $t>0$ 
in general. 
%
\end{example}
The Hamiltonian of the clock in the above example has 
an unbounded spectrum ${\bf R}$. This unbounded character must be 
satisfied in general. 
A result given by Hergerfeldt \cite{Hergerfeldt} can be directly applied to 
obtain the following theorem. 
\begin{theorem}
Spectrum of the clock Hamiltonian $H_c$ satisfying both Condition \ref{cond1} and Condition \ref{cond2} is unbounded 
when the system Hamiltonian $H_s$ and the interaction $V$ are bounded. 
\end{theorem} 
\begin{proof}
Suppose that $H_c$ is lower bounded. (An Upper bounded case can be treated similarly.) 
For an arbitrary $\Psi \in {\cal H}_c\otimes {\cal H}_s$ and an arbitrary state $\Omega\in {\cal H}_s$ we define  
\begin{eqnarray*}
f_{\Psi,\Omega}(t):=( \Psi,  V e^{-i \frac{H}{\hbar}t}
(\phi_c(0)\otimes \Omega)). 
\end{eqnarray*}
From Condition \ref{cond1}, this is vanishing for $t\leq 0$. 
As $H$ is lower bounded,  
$f_{\Psi,\Omega}(z):=\langle \Psi|Ve^{-i \frac{H_c}{\hbar}z}|\phi_c\otimes 
\Omega\rangle $ can be defined for $Im(z)\leq 0$ and is analytic 
for $Im(z)<0$. 
The Schwarz reflection principle concludes that $f_{\Psi,\Omega}$ can be 
extended to an analytic function on 
${\bf C}\setminus \{s|s > 0\}$. 
Because $f_{\Psi, \Omega}(t)=0$ on $t\leq 0$, it follows 
that $f_{\Psi,\Omega}(t)=0$ for $t\in {\bf R}$.  
That is, $Ve^{-i \frac{H}{\hbar}t} ( \phi_c(0)\otimes \Omega) =0$. 
\end{proof}
\section{time-energy uncertainty relation} 
In this section, we examine the behavior of the controlled system.  
Throughout this section, we consider the 
time evolution of 
$
 \Theta_0(0):=|\phi_c(0)\rangle \langle \phi_c(0)|
\otimes \rho_s(0). 
$ Its evolved state is denoted by $\Theta_0(t)=e^{-i\frac{H}{\hbar}t}\Theta_0(0)
e^{i\frac{H}{\hbar}t}$ whose restricted state on the system 
is written as $\rho_s(t):=\mbox{tr}_{{\cal H}_c}\Theta_0(t)$. 
This state $\rho_s(t)$ should be compared to  
the freely evolved state $\rho^0_s(t):=e^{-i\frac{H_s}{\hbar}t}
\rho_s(0)e^{i\frac{H_s}{\hbar}t}$. 
We discuss the extent to which the controlled state $\rho_s(t)$ can deviate from 
$\rho_s^0(t)$.
\par 
To quantify the deviation, the following two quantities are employed. 
The fidelity between two states $\rho_0$ and $\rho_1$is defined by 
$F(\rho_0,\rho_1):=\mbox{tr}(\sqrt{\rho_0^{1/2}\rho_1\rho_0^{1/2}})$. 
It satisfies $0\leq F(\rho_0,\rho_1)\leq 1$, and 
$F(\rho_0,\rho_1)=1$ holds if and only if $\rho_0=\rho_1$. 
The trace distance is another important quantity defined by 
$D(\rho_0,\rho_1)=\mbox{tr}(|\rho_0-\rho_1|)$. 
It satisfies the axioms for distance and $0\leq D(\rho_0,\rho_1)\leq 2$. 
 (See for example, \cite{Nielsen}.) 
\par 
As a quantity characterizing the time scale, 
\begin{eqnarray*}
\Delta H_c :=\sqrt{(\phi_c(0), H_c^2 \phi_c(0))-(\phi_c(0), H_c\phi_c(0))^2}
\end{eqnarray*}
plays a central role in the following theorems. 
It is known that this quantity characterizes the speed of time evolution for the isolated clock.
In fact, the Mandelstam-Tamm time-energy 
uncertainty relation \cite{Mandelstam-Tamm} is applied to 
show for $t\leq \frac{\pi \hbar}{2\Delta H_s}$ \cite{BuschTEUR, Pfeifer}
\begin{eqnarray}
|(\phi_c(t),\phi_c(0))|\geq \cos\left(
\frac{\Delta H_s t}{\hbar}
\right). 
\label{Mandelstam}
\end{eqnarray}
The deviation of the controlled state, measured by the fidelity, can be 
bounded by the following simple argument. 
\begin{theorem}
$F(\rho_s(t), \rho_s^0(t))$ represents the fidelity between $\rho_s(t)$ and 
the freely evolved state $\rho_s^0(t)$. 
It holds that for $t\leq \frac{\pi \hbar}{2\Delta H_c }$,   
\begin{eqnarray*}
\cos \left( \frac{\Delta H_c t}{\hbar}\right) \leq F(\rho_s(t), \rho_s^0(t)). 
\end{eqnarray*}
\end{theorem}
\begin{proof}
Let us consider two states $\Theta_0(0)$ and $\Theta_t(0)$ $(0\leq t\leq \frac{\pi \hbar}{2 \Delta H_c})$ defined by 
\begin{eqnarray*}
\Theta_0(0)&:=&|\phi_c(0)\rangle \langle \phi_c(0)|
\otimes \rho_s(0) 
\\
\Theta_t(0)&:=&|\phi_c(-t)\rangle \langle \phi_c(-t)|
\otimes \rho_s(0). 
\end{eqnarray*}
These states evolve with the Hamiltonian $H=H_c+H_s+V$. 
Let us denote the states at time $t$ by $\Theta_0(t)$ and $\Theta_t(t)$. 
While $\Theta_0(t)$ may have a complicated form, $\Theta_t(t)$ has 
a simple form,
\begin{eqnarray*}
\Theta_t(t)=\phi_c(0)\otimes \rho_s^0(t), 
\end{eqnarray*}
where we used 
Lemma \ref{lemma1}.
Because the fidelity between two states is invariant 
under unitary evolution \cite{Nielsen}, it follows that 
\begin{eqnarray*}
F(\Theta_0(0), \Theta_t(0))
=F(\Theta_0(t), \Theta_t(t)). 
\end{eqnarray*}
The left-hand side of the above equation becomes 
\begin{eqnarray*}
F(\Theta_0(0), \Theta_t(0))
=|(\phi_c(0), \phi_c(-t))|
\end{eqnarray*}
and the right-hand side is bounded as 
\begin{eqnarray*}
F(\Theta_0(t), \Theta_t(t))\leq F(\rho_s(t), \rho_s^0(t)), 
\end{eqnarray*}
where we utilized the fact that the fidelity decreases for restricted states \cite{Nielsen}. 
Thus it holds that 
\begin{eqnarray*}
|(\phi_c(0),\phi_c(-t))|\leq F(\rho_s(t),\rho_s^0(t)). 
\end{eqnarray*} 
The left-hand side of this inequality represents the speed of 
time evolution of the clock and is bounded as (\ref{Mandelstam}).
It concludes the proof. 
\end{proof}
An inequality $\cos\left(\frac{\Delta H_c t}{\hbar}\right)
\geq 1- \frac{2 \Delta H_c t}{\pi \hbar}$ is applied to the above theorem 
to obtain a weaker but simpler form.  
\begin{corollary}
For $t\leq \frac{\pi \hbar}{2\Delta H_c}$, it holds that 
\begin{eqnarray*}
t\cdot \Delta H_c \geq \frac{\pi \hbar}{2}\left(1-F(\rho_s(t), \rho_s^0(t)\right). 
\end{eqnarray*}
\end{corollary}
In the following, we treat the trace distance as a quantity measuring the deviation. 
Applying an inequality 
$D(\rho_0,\rho_1)\leq 2 \sqrt{1-F(\rho_0,\rho_1)^2}$ \cite{Nielsen} to the above theorem, 
we can obtain 
$
\cos^2 \left(\frac{\Delta H_c t}{\hbar}\right)\leq 1-\frac{D(\rho_s(t),\rho^0_s(t))^2}{4}.
$
This inequality can be improved as follows. 
\begin{theorem}
For $t\leq \frac{\pi \hbar}{2\Delta H_c}$, it holds that 
\begin{eqnarray*}
\cos^2\left(\frac{\Delta H_c t}{\hbar}
\right)\leq 1-\frac{D(\rho_s(t), \rho_s^0(t))}{2}.
\end{eqnarray*}
\end{theorem}
\begin{proof}
For $0\leq t\leq \frac{\pi \hbar}{2\Delta H_s}$, we consider two states 
\begin{eqnarray*}
\Theta_0(0)&:=&|\phi_c(0)\rangle \langle \phi_c(0)|
\otimes \rho_s(0) 
\\
\Theta_t(0)&:=&|\phi_c(-t)\rangle \langle \phi_c(-t)|
\otimes e^{i\frac{H_s}{\hbar}t}|\xi\rangle \langle \xi|e^{-i\frac{H_s}{\hbar}t}, 
\end{eqnarray*}
where $|\xi\rangle$ is an arbitrary state of the system. 
The time-evolved states are denoted by $\Theta_0(t)$ and $\Theta_t(t)$,  
respectively. 
$\Theta_t(t)$ can be written as 
$
\Theta_t(t)=|\phi_c(0)\rangle \langle \phi_c(0)|\otimes |\xi\rangle \langle \xi|, 
$
where we used Lemma \ref{lemma1}.
Invariance of fidelity under unitary operations derives 
\begin{eqnarray*}
F(\Theta_0(0),\Theta_t(0))
=F(\Theta_0(t),\Theta_t(t)).
\end{eqnarray*}
The left-hand side of this equation can be written as
\begin{eqnarray*}
F(\Theta_0(0),\Theta_t(0))
=| (\phi_c(0), \phi_c(-t))| 
(e^{i\frac{H_s}{\hbar}t}\xi, \rho_s(0) e^{i\frac{H_s}{\hbar}t}\xi)^{1/2}
=|(\phi_c(t),\phi_c(0))|(\xi,\rho_s^0(t)\xi)^{1/2},
\end{eqnarray*}
where $\rho_s^0(t):=e^{-i\frac{H_s}{\hbar}t}\rho_s(t)e^{i\frac{H_s}{\hbar}t}$ 
is the freely evolved state of the system. 
The right-hand side can be 
bounded as 
\begin{eqnarray*}
F(\Theta_0(t),\Theta_t(t))
\leq (\xi,\rho_s(t)\xi)^{1/2}. 
\end{eqnarray*}
Thus we obtain 
\begin{eqnarray*}
|(\phi_c(t),\phi_c(0))|^2(\xi,\rho_s^0(t)\xi)
\leq (\xi,\rho_s(t)\xi). 
\end{eqnarray*} 
Because $|(\phi_c(t),\phi_c(0))|\geq \cos \left( \frac{\Delta H_c t}{\hbar} \right)$ holds 
for $t\leq \frac{\pi \hbar}{2\Delta H_c}$, 
it follows that 
\begin{eqnarray}
\cos^2 \left(\frac{\Delta H_c t}{\hbar}\right)(\xi,\rho_s^0(t)\xi)
\leq (\xi,\rho_s(t)\xi). 
\label{ineq1}
\end{eqnarray}
This can be deformed as 
\begin{eqnarray}
\cos^2 \left(\frac{\Delta H_c t}{\hbar}\right) (\xi,\rho_s^0(t)-\rho_s(t) \xi)
\leq 
\left(1- \cos^2 \left(\frac{\Delta H_c t}{\hbar}\right)
\right)
(\xi,\rho_s(t)\xi).
\label{tochu} 
\end{eqnarray}
We denote by $E^+$ a projection operator onto the subspace 
$\{x| (\rho_s^0(t)-\rho_s(t)) x =\lambda x, \lambda \geq 0\}$.  
Because $\xi$ is arbitrary, we consider an orthonormalized family $\{\xi\}$ to 
satisfy $\sum_{\xi}|\xi\rangle \langle \xi|=E^+$. Taking the 
summation of eq. (\ref{tochu}) with respect to 
this family, we obtain 
\begin{eqnarray*}
\cos^2 \left(\frac{\Delta H_c t}{\hbar}\right) 
\frac{D(\rho_s(t), \rho_s^0(t))}{2}&\leq& \left(1- \cos^2 \left(\frac{\Delta H_c t}{\hbar}\right)
\right) \mbox{tr}(\rho_s(t)E^+)
\\
&\leq & 
 \left(1- \cos^2 \left(\frac{\Delta H_c t}{\hbar}\right)\right)
\left(1-\frac{D(\rho_s(t),\rho_s^0(t))}{2}\right), 
\end{eqnarray*}
where we used the relationship 
$\mbox{tr}(\rho_s(t)E^+)=\mbox{tr}(\rho_s^0(t)E^+)-\frac{D(\rho_s^0(t),\rho_s(t))}{2}$ 
and $\mbox{tr}(\rho_s^0(t)E^+)\leq 1$. 
The above inequality is equivalent to $
\cos^2\left(\frac{\Delta H_c t}{\hbar}
\right)\leq 1-\frac{D(\rho_s(t), \rho_s^0(t))}{2}.$
\end{proof}
According to the above theorems, we can see that at least $\frac{\pi \hbar}{2\Delta H_c}$ is required to 
deviate the state perfectly from the original dynamics. 
It should be noted that this time scale is determined only by the energy fluctuation of the 
clock. 
\begin{example}
(Einstein's photon box - Time-Energy uncertainty relation for event) 
It is well-known that Einstein repeatedly challenged the consistency of quantum physics. One of the 
famous Gedanken experiment he proposed is the so-called photon box experiment. 
Although there have been debates \cite{Dieks} on what Einstein actually intended with this experiment, 
in this paper, 
we just describe the experiment and analyze it as follows.
\par 
Suppose there exists a box furnished with a small shutter that is controlled by a clock inside the box. 
The clock-shutter mechanism is tuned so as to let the shutter open in a 
very short time interval $t$. 
During this interval, a photon is supposed to escape from the box to the outside. 
The energy of the photon is determined by measuring the energy of the box after the 
photon has escaped. 
Thus obtained box energy is subtracted from the box energy before the opening period to 
 obtain the energy of photon. Einstein argued that the photon energy 
can be determined in an arbitrary precision by such a procedure. 
\par
While Bohr replied to this challenge by ingeniously combining general relativity and 
the position-momentum uncertainty relation, his argument has 
left many researchers unsatisfied because  
it strongly depends on the specific method of measurement and involves gravity. 
Busch \cite{Busch1} offered  a clear explanation that makes no assumptions 
regarding  
the method of measurement. We describe it with an emphasis on the applicability of 
our theorem. 
\par
The entire system consists of the box and the outside. 
The box is furnished with a clock and a shutter. The box also contains
photons. The outside is an arbitrary system. 
The box is regarded as a clock system in our context. 
During a certain time interval, the box changes the 
state of the outside.  
To discuss without the state change of the outside without ambiguity, 
the state after the interaction must be completely 
distinguishable from the unperturbed state. 
For example, a photon is added to the vacuum. 
Thus, $F(\rho^0_s(t),\rho_s(t))=0$ should hold. 
This requires that the energy of the box before the interaction 
should have energy fluctuation $\Delta H_c$ satisfying 
$t\cdot \Delta H_c \geq \frac{\pi \hbar}{2}$. 
Thus the energy of the photon cannot be measured 
beyond the accuracy determined by this $\Delta H_c$. 
\par
One may read this scenario as follows. 
Suppose that there exists a measurement apparatus in the 
outside. To bring about a notable change of this apparatus after time $t$,  
the box must have energy fluctuation $\Delta E$ determined by $t\cdot \Delta E 
\geq \frac{\pi \hbar}{2}$. 
If we regard this notable change of 
the apparatus as an event, we may interpret the 
result as a version of event time-energy uncertainty relation. 
\end{example}
According to eq. (\ref{ineq1}) in the proof of the theorem, 
$\mbox{supp}\rho_s^0(t)\subset \mbox{supp}\rho_s(t)$ must be satisfied for $t\leq \frac{\pi \hbar}{2\Delta H_c}$. 
From this, we can conclude that the dynamics of the system cannot be unitary unless it is trivial. 
In fact, for pure $\rho^0_s(t)=|\phi \rangle \langle \phi |$, $\rho_s(t)$ cannot be pure unless 
$\rho_s(t)=|\phi\rangle \langle \phi|$ holds. 
Thus,  
the phenomenological treatment of control process using a time-dependent Hamiltonian is not 
rigorously valid for the short time range $t\leq \frac{\pi \hbar}{2 \Delta H_c}$. 
\par
The above discussion shows that the clock is also disturbed for a short time range. 
In fact when a pure state of the system becomes mixed, the whole state must be an entangled state. 
That is, the state of the clock is also changed to a mixed state. 
The following example shows that this back action on the clock may no longer 
exist 
 for $t\geq \frac{\pi \hbar}{2 \Delta H_c}$. 
\begin{example}
Using the clock and system in Example \ref{ex1},  
let us consider the time evolution of the state $|\phi(0)\rangle \otimes ( c_0 |0\rangle +c_1 |1\rangle )$.  
If $g(x)$ is chosen to satisfy $\int^{\delta}_0 dx g(x)=\pi \hbar$;  
for $t \geq \delta +\Delta$,  
the state evolves as 
\begin{eqnarray*}
|\phi_c(t)\rangle \otimes( c_0 |0\rangle - c_1 |1\rangle ). 
\end{eqnarray*}
\end{example}
In this example, the interaction leaves no trace on the clock after the interaction, 
whereas the system evolves nontrivially. 
Thus we can conclude that the back action to the clock does not take 
necessarily place for 
large $t$.
\section{discussion}
In this paper, 
a definition of a quantum clock working as a control device was introduced, 
and its property was 
examined. We proved that a clock satisfying the definition has an unbounded Hamiltonian spectrum. 
The fidelity and trace distance between perturbed and unperturbed 
states characterize the state change of the controlled system.  
We presented uncertainty relations that relate the time duration of the process and the energy of the clock to 
the magnitude of the state change. 
According to these relations, time $\frac{\pi \hbar}{2\Delta H_c}$ is required to 
deviate the state of the system completely. 
In addition we have shown that for a short time $t <\frac{\hbar \pi}{2\Delta H_c}$ 
a pure state is perturbed to a mixed state unless it is unperturbed. 
While the clock is perturbed for a short time, this back action on the clock may no longer 
exist 
 for large $t$. 
\par
While our definition and the time-energy relation are simple, 
considerable work remains from the realistic perspective. 
Because the definition is so strict, 
the Hamiltonian of the clock must have an 
unbounded spectrum. Although such an unbounded 
Hamiltonian is common in treating infinite systems, a realistic clock should be a finite system, and 
hence, the nonlocality of time will be critical. The magnitude of such an effect should be examined carefully. 
 Controlling quantum states plays an important role in various fields. In chemistry, 
to accelerate chemical reaction various methods to 
control quantum states have been developed.  
In quantum information, the 
accuracy and speed of control have a direct effect on the quality of the quantum computer. 
It would be interesting to study possible effects of our argument 
in such applications. 


\begin{thebibliography}{9}

\bibitem{Wigner}H.~Salecker and E.~P.~Wigner, Phys. Rev. {\bf 109}, 571 (1958). 

\bibitem{Peres}A.~Peres, Am. J. Phys. {\bf 48}(7), 552 (1980). 

\bibitem{Hilgevoord2}J.~Hilgevoord, Am.~J.~Phys. {\bf 70}(3), 301 (2002).





\bibitem{Busch1}P.~Busch, Found. Phys. {\bf 20}, 1 (1990).  

\bibitem{BuschTEUR}P.~Busch, in  
{\em Time in Quantum Mechanics}, eds. J. G. Muga, R. Sala Mayato, I.L. Egusquiza, Springer-Verlag, Berlin 2nd ed. (2008). 


\bibitem{Hilgevoord1}J.~Hilgevoord, Am. J. Phys. {\bf 66}, 396 (1998). 

\bibitem{AReznik}Y.~Aharonov and B.~Reznik, 
Phys. Rev. Lett. {\bf 84}, 1368 (2000). 
\bibitem{CReznik}A.~Casher and B.~Reznik, Phys. Rev. A {\bf 62}, 042104 (2000). 


\bibitem{mixed}
Although the states of the clock may not be pure states, our treatment is 
general because  
the mixed states can be treated as pure states by enlarging the Hilbert space 
with a purification technique. 


\bibitem{Hergerfeldt}G.~C.~Hergerfeldt, Phys. Rev. Lett. {\bf 72} 596 (1994). 

\bibitem{Nielsen}M.~Nielsen and I.~Chuang, {\em Quantum Computation and Quantum Information}, Cambridge University Press, (2000).

\bibitem{Mandelstam-Tamm}L.~Mandelstam and I.~G.~Tamm, J. Phys. U.S.S.R. {\bf 9}, 
249 (1945). 

\bibitem{Pfeifer}P.~Pfeifer, Phys. Rev. Lett. {\bf 70}, 3365 (1993). 


\bibitem{Dieks}D.~Dieks and S.~Lam, Am. J. Phys. {\bf 76}, 838 (2008). 

\end{thebibliography}
\end{document}